\documentclass[twocolumn,preprintnumbers,superscriptaddress,nofootinbib]{revtex4-2}
\usepackage{graphicx}
\usepackage{dcolumn}
\usepackage{bm}
\usepackage{makecell}
\usepackage{braket}
\usepackage{lipsum}
\usepackage{ulem}
\usepackage{amsmath,amssymb} 
\usepackage[export]{adjustbox}
\usepackage{float}
\usepackage{caption}
\usepackage{subcaption}
\usepackage{enumitem}
\usepackage{xcolor}
\usepackage{graphicx}
\usepackage{csquotes}

\usepackage[colorlinks=true, linkcolor=red, urlcolor=blue, citecolor=blue]{hyperref}
\usepackage{microtype}
\usepackage{orcidlink}
\DeclareUnicodeCharacter{2212}{-}

\usepackage[a4paper,top=2.54cm,bottom=2.54cm,left=2.54cm,right=2.54cm,marginparwidth=1.75cm]{geometry}

\makeatletter
\DeclareRobustCommand{\HI}{%
  \mbox{H\check@mathfonts\fontsize\sf@size\z@\selectfont I}%
}
\makeatother

\begin{document}
        \title{Observations of the 21 cm \HI{} Line from the Milky Way galaxy using Pyramidal Horn Radio Telescope}

        \author{Kaustav Bhattacharjee\,\orcidlink{0000-0001-7437-3916}}
        \email[K. Bhattacharjee: ]{kaustav\_b@ph.iitr.ac.in}
        \affiliation{Department of Physics, Indian Institute of Technology Roorkee, Roorkee, India}
        \affiliation{Raman Research Institute, Bengaluru, India}
        
        \author{Himanshu Grover\,\orcidlink{0009-0004-1150-6151}}
        \affiliation{Department of Physics, Indian Institute of Technology Roorkee, Roorkee, India}
        
        \author{P. Arumugam\,\orcidlink{0000-0001-9624-8024}}
        \affiliation{Department of Physics, Indian Institute of Technology Roorkee, Roorkee, India}


\begin{abstract}
We present the design, implementation, and operation of a pyramidal horn radio telescope built for detecting the Galactic 21 cm neutral hydrogen line emission. The system employs an SDR-based pipeline to obtain drift-scan observations, which were calibrated and processed to generate \HI{} sky maps, a Galactic rotation curve and spiral arm features. This demonstrates that this low-cost system is effective both for educational purposes and scientific exploration of Galactic structure at radio frequencies.
\end{abstract}

\maketitle

\section{Introduction}

The hyperfine transition line of neutral hydrogen (\HI{}) provides a powerful tool for probing the structure and dynamics of the Milky Way and other galaxies. Spin-spin coupling between the electron and proton of neutral hydrogen atom breaks the spin degeneracy of its $1s$ ground state and produces two hyperfine levels with an energy gap $\Delta E$ given by:
\begin{equation}
    \Delta E = \frac{4 g_p \hbar^4}{3 m_p m_e^2 c^2 a_0^4} = 5.874 \,\mu\text{eV}, 
\end{equation}
where all symbols have their usual physical meaning (see \citeauthor{griffiths2018quantum} \citeyear{griffiths2018quantum}). The frequency $\nu$ of the photon emitted in a transition from the triplet state to the singlet state, using Planck's formula relating energy and frequency, is
\begin{equation}
    \nu = \frac{\Delta E}{h} = 1420.405\,\text{MHz}.
\end{equation}
Dutch astronomer Hendrik van de Hulst predicted the existence of this radio spectral line in 1944 (\citeauthor{vandeHulst1945} \citeyear{vandeHulst1945}). Jan Oort recognised that, due to its long wavelength, such a spectral line would allow astronomers to probe the structure and differential rotation of the Galaxy through the extensive clouds of dust in the Galactic plane that absorb all visible light within merely a few thousand light years (\citeauthor{NRAO_21cm_prediction} \citeyear{NRAO_21cm_prediction}).
 Since its first detection in March of 1951 by Harold Ewen and Edward Purcell (\citeauthor{ewen1951observation} \citeyear{ewen1951observation}) using a horn antenna at the Lyman Laboratory of Harvard University, the 21 cm hydrogen line has become a cornerstone of astrophysical research. 
 This emission line is a key observational feature in radio astronomy, providing critical insights into Galactic rotation curves and spiral arm structures (\citeauthor{Hulst1954} \citeyear{Hulst1954}), and the distribution of interstellar matter.

\begin{figure}
    \centering
    \includegraphics[width=0.48\textwidth]{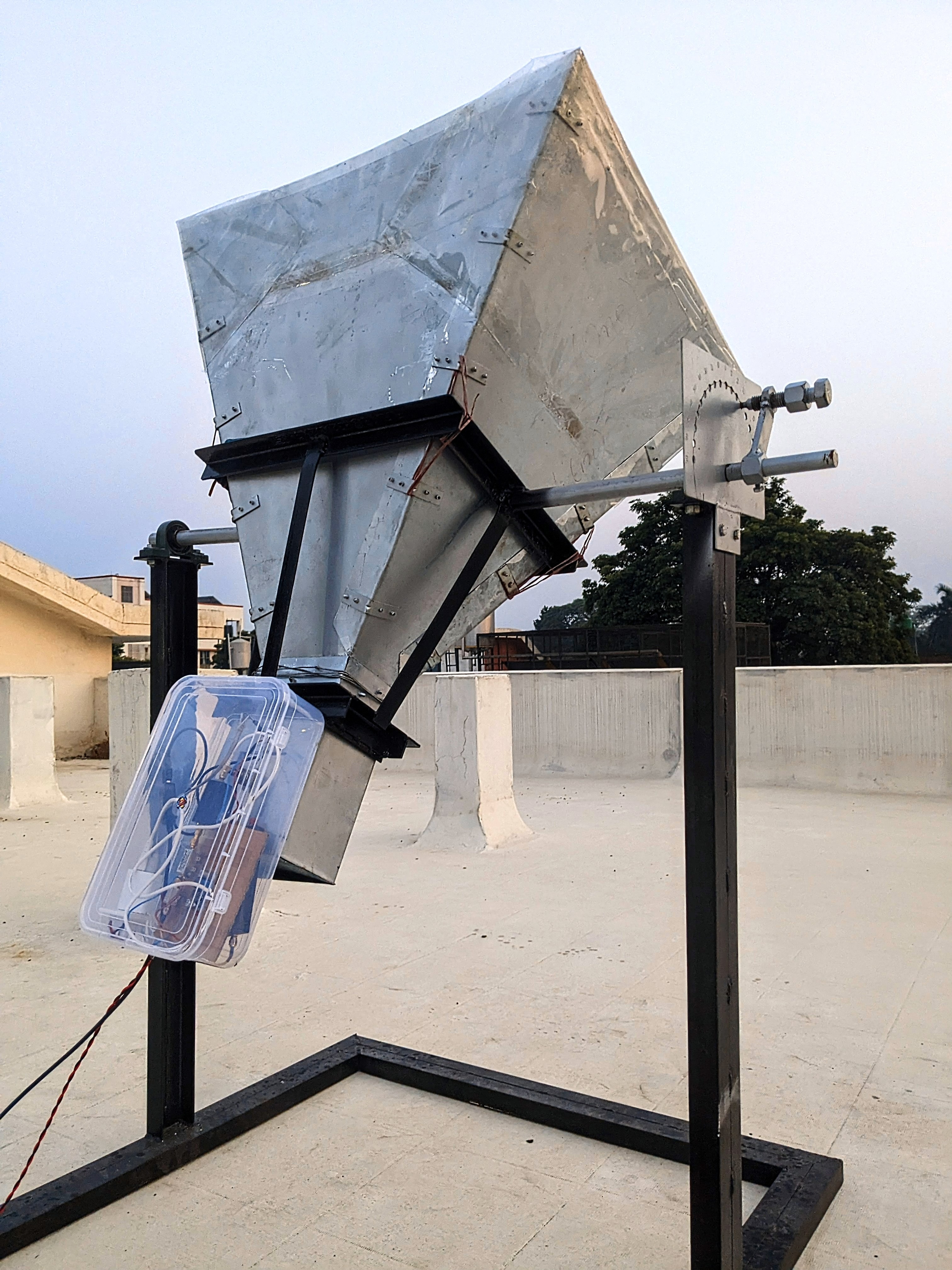}
    \caption{Pyramidal Horn Radio Telescope.}
    \label{fig:horn}
\end{figure}

In recent years, there has been a growing interest in developing accessible and low-cost radio telescope systems to enable student and amateur participation in observational radio astronomy; some of these include instruments such as the BHARAT dual-mode conical horn telescope demonstrated by \citeauthor{Mhaske2022} (\citeyear{Mhaske2022}), the 91cm x 81cm aperture pyramidal horn telescope made by \citeauthor{langston2017light} (\citeyear{langston2017light}), and the 70 cm x 60 cm aperture pyramidal horn designed by \citeauthor{Pandian2022-pd} (\citeyear{Pandian2022-pd}). We present the design and implementation of a pyramidal horn radio telescope tailored for detecting the 21 cm \HI{} line, developed as part of the activities of the Radio Telescope Project\footnote{\url{https://iitr.ac.in/rtp}} of the Physics and Astronomy Club\footnote{\url{https://paac.iitr.ac.in/}} at IIT Roorkee. Our horn design emphasises mechanical simplicity and effective signal collection.

The telescope system described here uses a pyramidal horn antenna optimised for operation at 1.42 GHz, combined with a low-noise receiver chain and software-defined radio (SDR) backend. The complete setup is shown in Figure~\ref{fig:horn}. The primary goal was to make a functional platform capable of detecting Galactic \HI{} emission with sufficient sensitivity and resolution to allow the study of the Milky Way’s rotation curve. The outline of the paper is as follows: Section~\ref{Construction of Horn and Waveguide} discusses the construction of the horn and waveguide. Section~\ref{The Radio Frequency Signal Chain} describes the analogue electronic components and the SDR used in the setup. Section~\ref{Simulation of Antenna Performance} then presents the electromagnetic simulations and the characterisation calculations of the antenna performance. Section~\ref{Software Architecture} discusses the GNU Radio flowgraph design. The observational strategy is presented in Section~\ref{Observational Strategy}, covering the scanning and calibration procedures. Section~\ref{Data Analysis} outlines the data analysis methodology. Finally, Section~\ref{Results and Discussion} presents the results of the observations, which include the processed Galactic maps, the derived velocity curve, and the identification of spiral arms.

\begin{figure*}
    \centering

    \begin{subfigure}{0.24\textwidth}
        \centering
        \includegraphics[width=\linewidth]{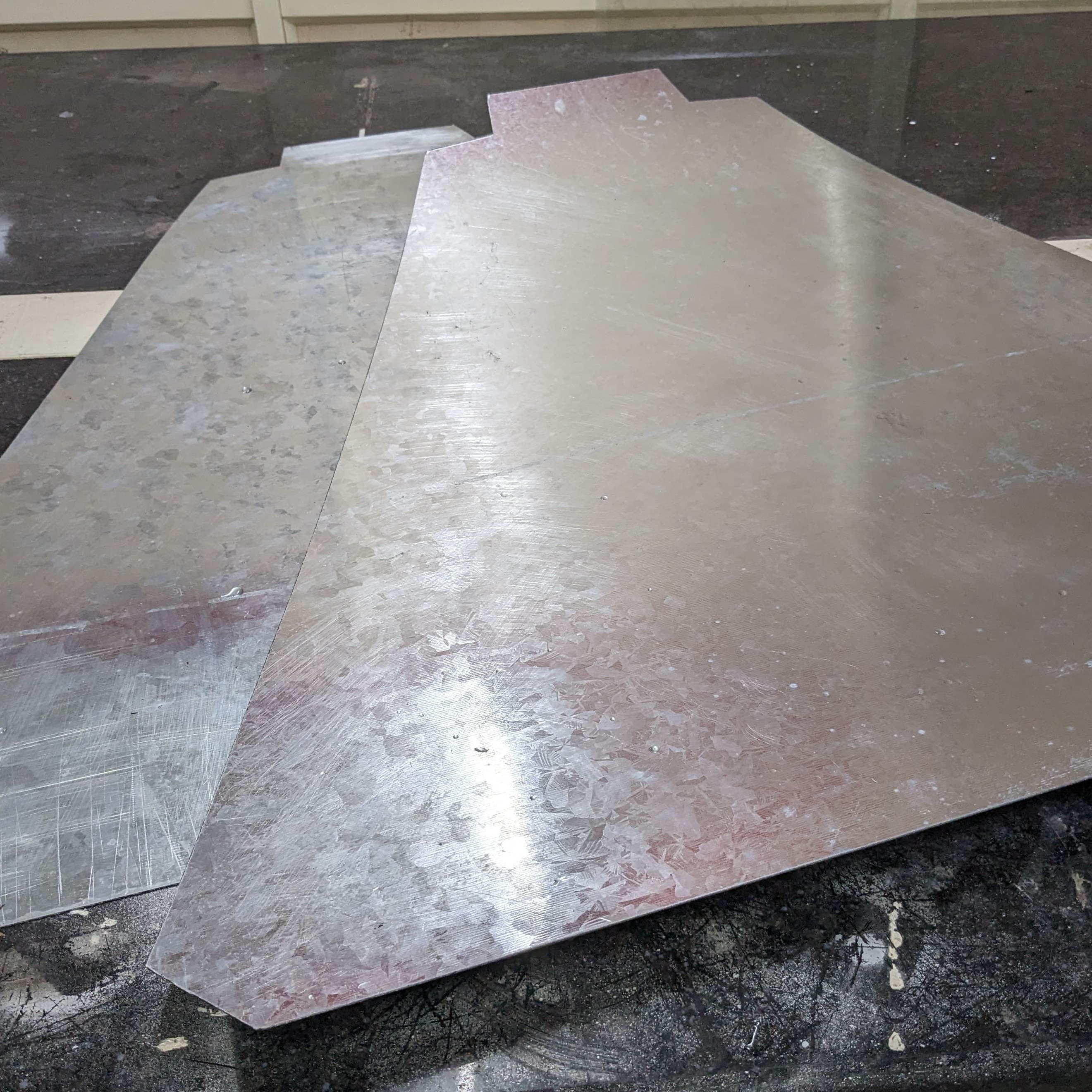}
        \caption{Laser-cut petals of the horn}
        \label{Laser-cut petals of the horn}
    \end{subfigure}
    \hfill
    \begin{subfigure}{0.24\textwidth}
        \centering
        \includegraphics[width=\linewidth]{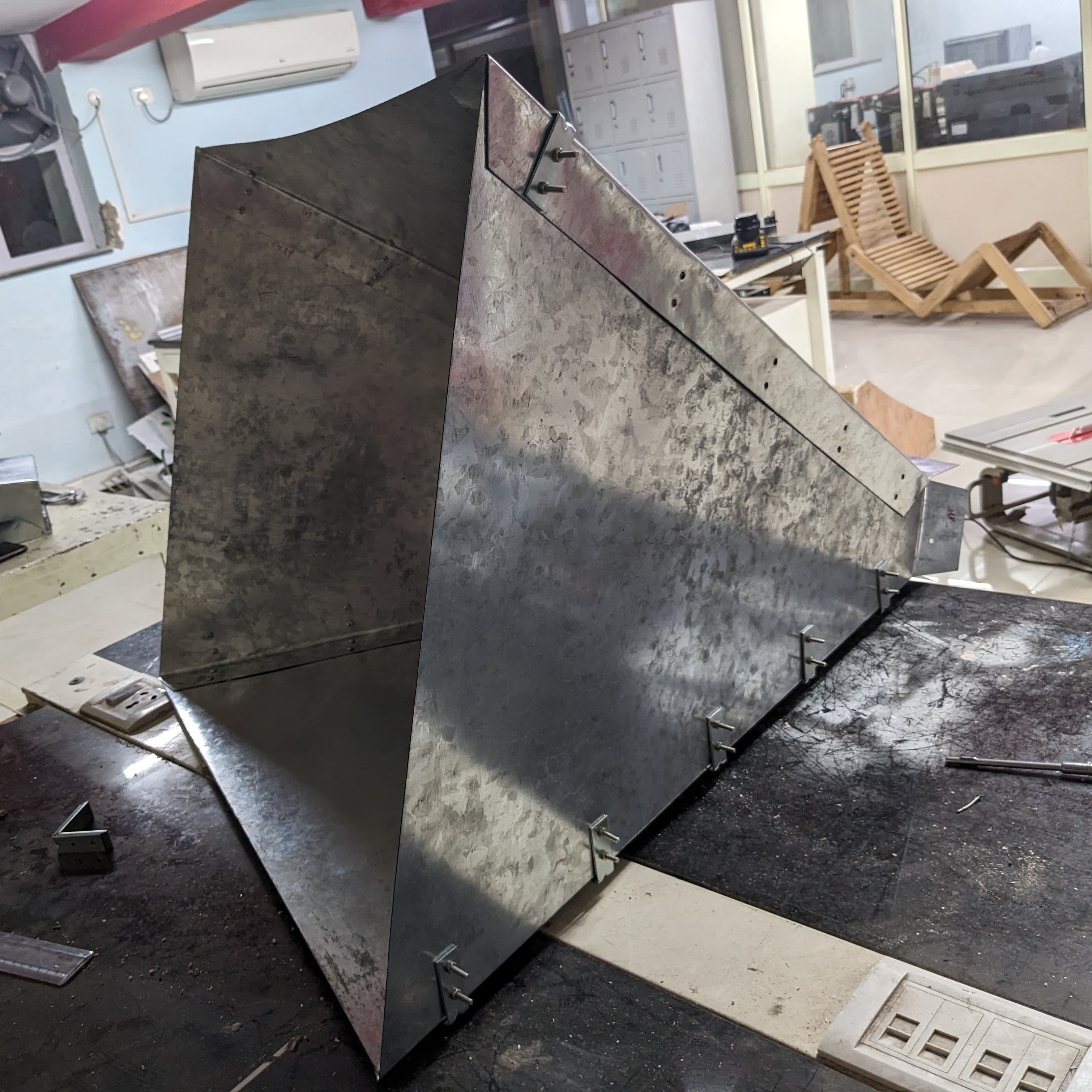}
        \caption{Folded and joined petals}
        \label{Folded and joined petals}
    \end{subfigure}
    \hfill
    \begin{subfigure}{0.24\textwidth}
        \centering
        \includegraphics[width=\linewidth]{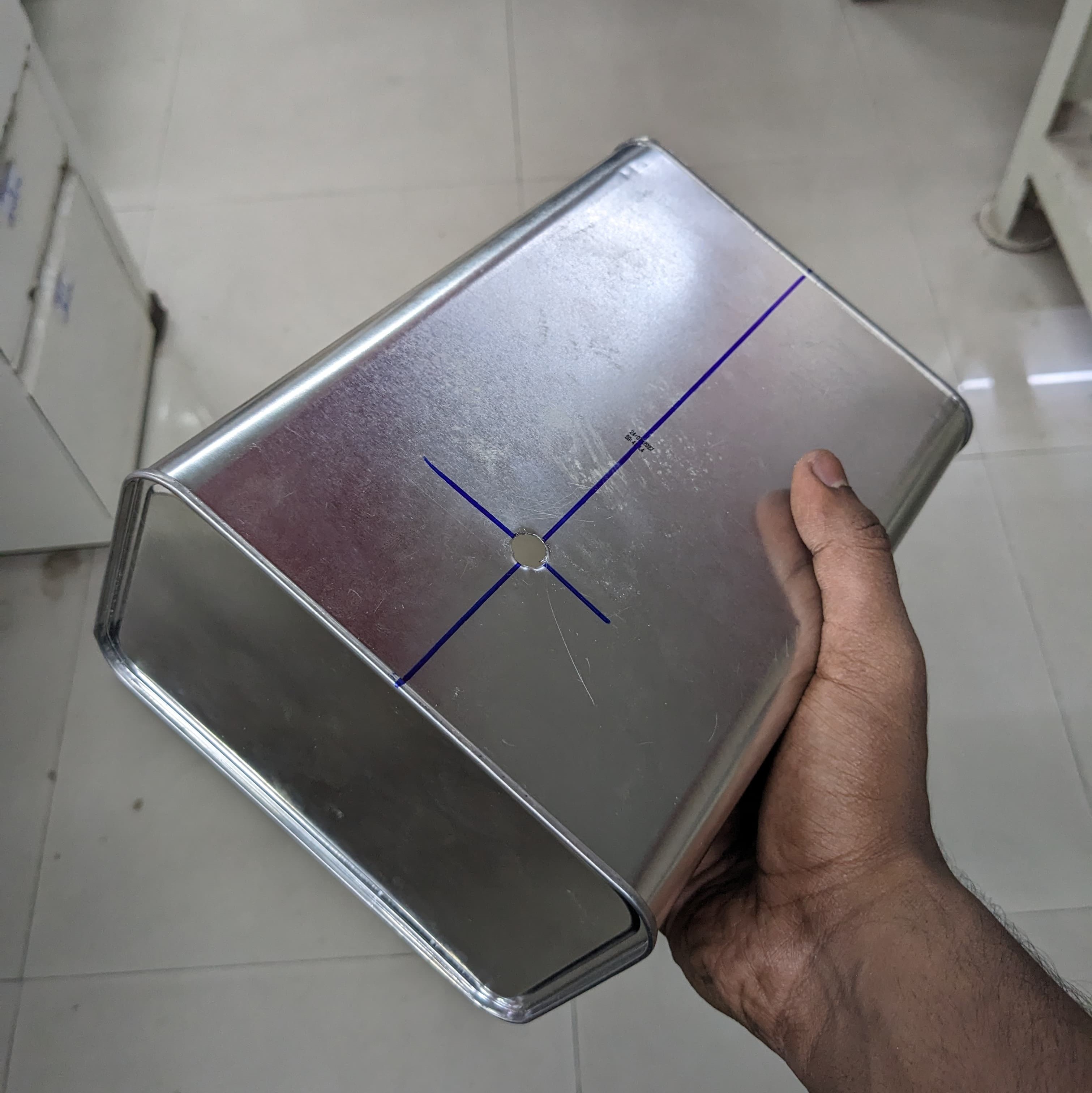}
        \caption{Waveguide with drilled hole for the probe}
        \label{Waveguide with drilled hole for the probe}
    \end{subfigure}
    \hfill
    \begin{subfigure}{0.24\textwidth}
        \centering
        \includegraphics[width=\linewidth]{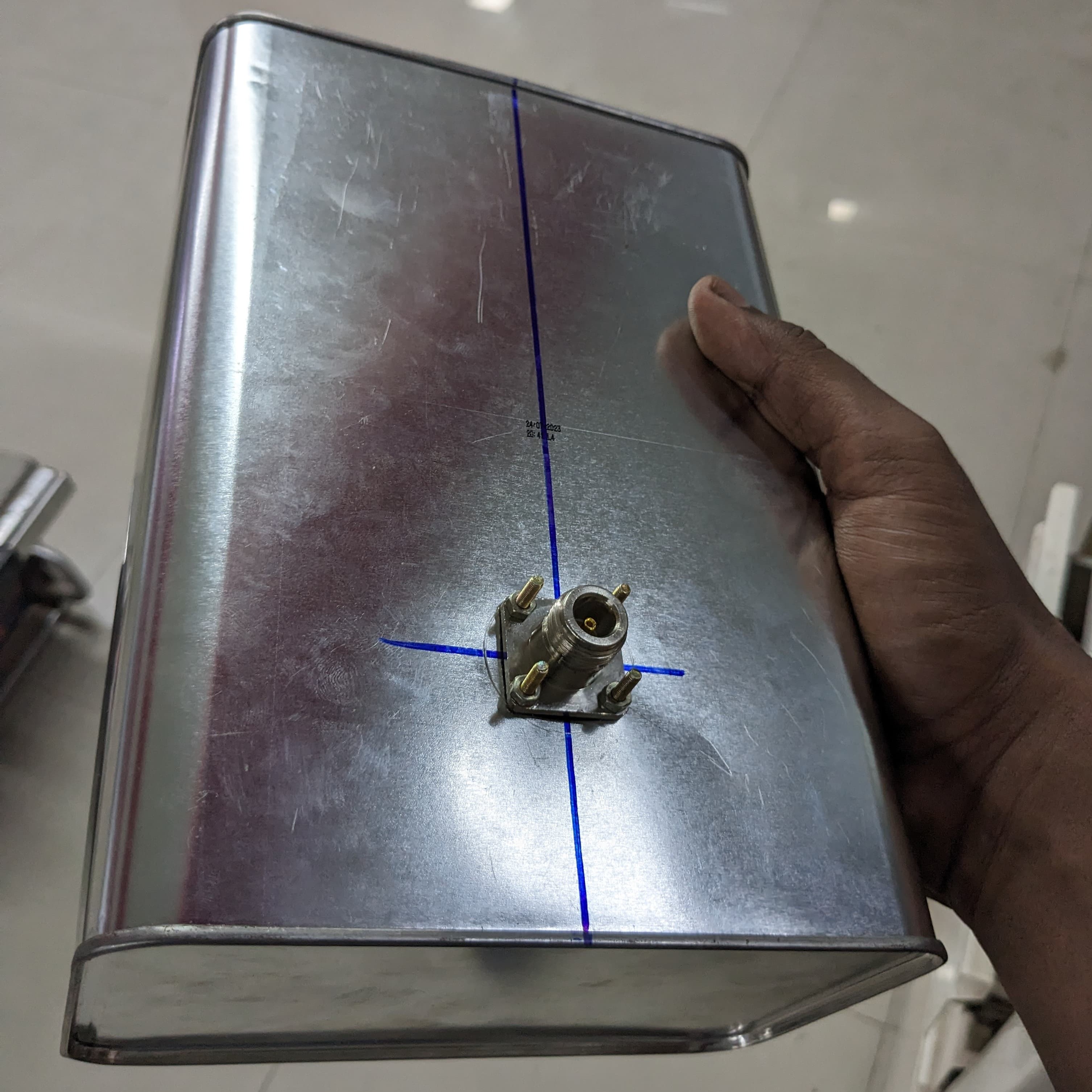}
        \caption{Probe attached to the waveguide}
        \label{Probe attached to the waveguide}
    \end{subfigure}
    \caption{Construction of the feedhorn assembly.}
    \label{fig:feedhorn_steps}
\end{figure*}

\section{Construction of Horn and Waveguide}
\label{Construction of Horn and Waveguide}

The feedhorn was constructed following the pyramidal design of the Digital Signal Processing in Radio Astronomy (DSPIRA) project  (\citeauthor{DSPIRA2025} \citeyear{DSPIRA2025}). The aperture dimensions were 75 cm × 60 cm, sufficiently large to allow passage through most doorways. The fabrication process is summarised in Figure~\ref{fig:feedhorn_steps}. The horn body was fabricated from 1 mm thick stainless steel sheets, which were laser-cut into trapezoidal petals with flared strips along the edges to allow assembly, as seen in Figure~\ref{Laser-cut petals of the horn}. The flares were bent to right angles relative to the petals by hammering them gently, after which the petals were aligned, drilled, and riveted together to form the horn structure, as seen in Figure~\ref{Folded and joined petals}.

At the narrow end of the horn, a stainless steel enclosure was similarly riveted to house the waveguide. A commercially available F-style 1-gallon rectangular paint thinner can was used as the waveguide cavity, chosen for its suitable dimensions of $16.8 \times 10.2 \times 24.1$~cm$^3$. The top portion of the can, including the handle and spout, was removed, and a feed port was prepared by drilling a hole in the sidewall 5.25 cm from the closed end, as seen in Figure~\ref{Waveguide with drilled hole for the probe}. An N-type female 4-hole panel mount connector, seen in Figure~\ref{Probe attached to the waveguide}, was screwed to the waveguide wall, with a 5.25 cm copper wire soldered to its central conductor and extending inside as the probe. The can was snugly fitted into the enclosure, allowing for easy insertion, removal, or replacement as required.

\begin{figure}
    \centering
    \begin{subfigure}{0.48\textwidth}
        \centering
        \includegraphics[width=\linewidth]{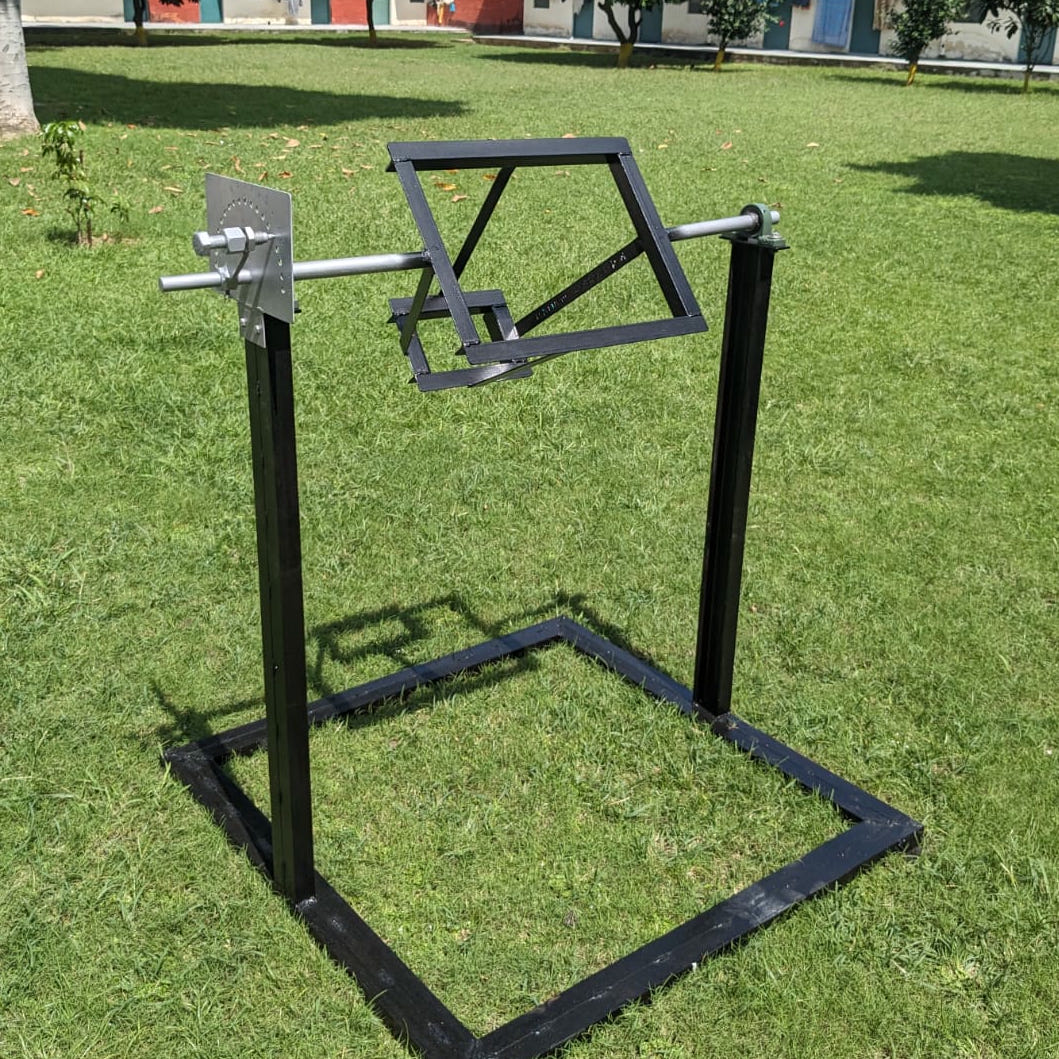}
        \caption{Support stand without Feedhorn}
        \label{only_stand}
    \end{subfigure}%
    \hfill
    \begin{subfigure}{0.48\textwidth}
        \centering
        \includegraphics[width=\linewidth]{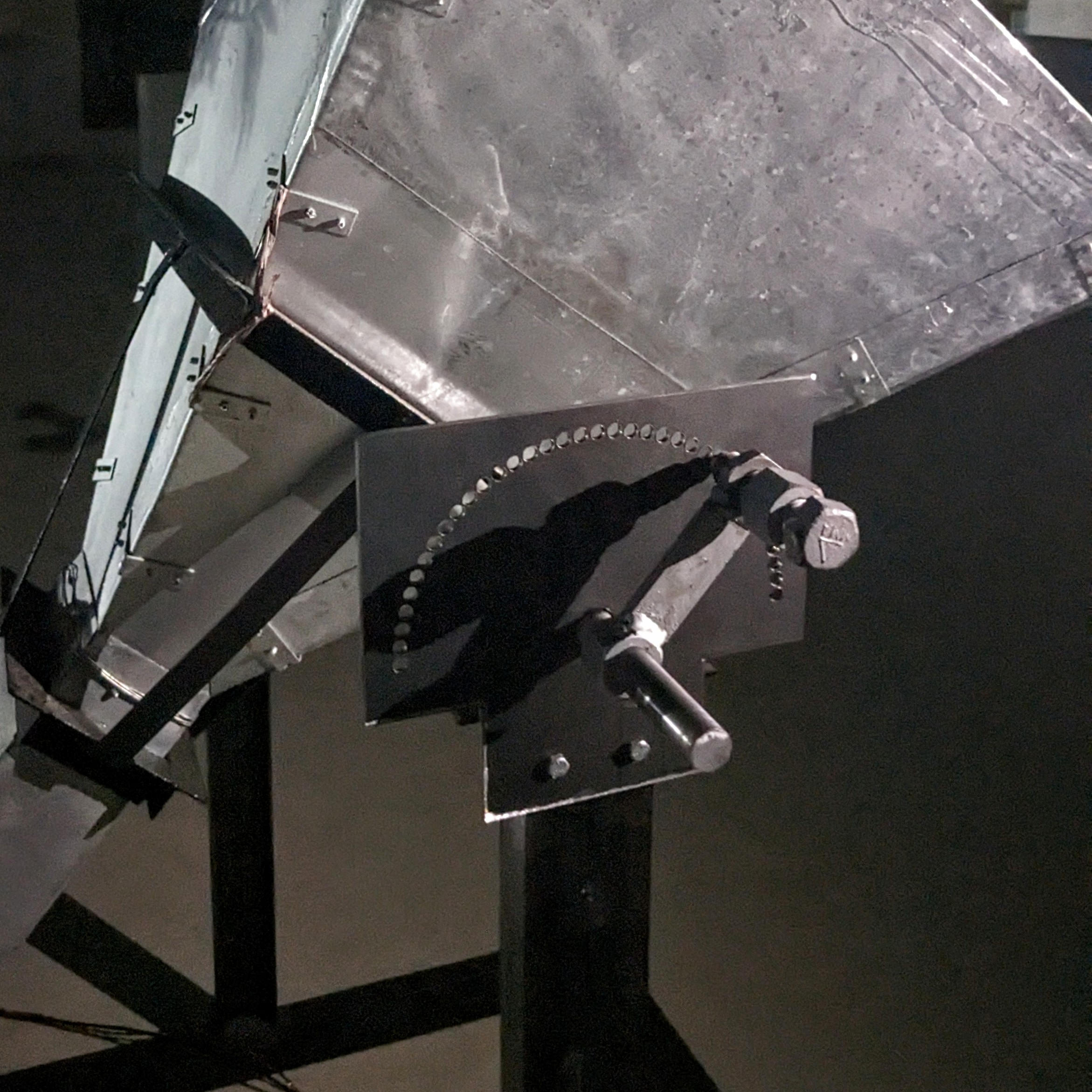}
        \caption{The elevation locking mechanism}
        \label{locking_mech}
    \end{subfigure}

    \caption{The support mechanism.}
    \label{fig:stand}
\end{figure}

The interior surfaces of both the horn and the waveguide were lined with aluminium foil tape to improve conductivity. Care was taken to ensure smooth, continuous coverage without gaps.
A support stand was manufactured to hold and orient the feedhorn assembly, shown in Figure~\ref{fig:stand}. The frame, seen without the horn in Figure~\ref{only_stand}, was constructed using heavy metal tube welded into a large square base for stability. Two vertical columns were fixed to the base, supporting a horizontal axle that allowed rotation of the mounted feedhorn in the elevation plane. An inner frame, composed of angled iron brackets, was welded to the axle. The feedhorn assembly was fixed in this inner frame by using cables. One end of the axle was equipped with a manual locking mechanism to secure the desired elevation angle. This locking mechanism comprised a steel bar ending in a cylindrical knob with a threaded hole. A locking screw passed through this knob and could be tightened into a perforated steel plate fixed to the outer frame. The plate, seen clearly in Figure~\ref{locking_mech}, featured an array of holes spaced 5 degrees apart distributed along a semicircular arc, corresponding to different elevation angles. By aligning the screw with the desired hole and tightening it, the feedhorn's elevation could be fixed securely, preventing any unintended motion during observation.

\section{The Radio Frequency Signal Chain}
\label{The Radio Frequency Signal Chain}
\begin{figure*}[ht!]
    \centering
    \includegraphics[width=\textwidth]{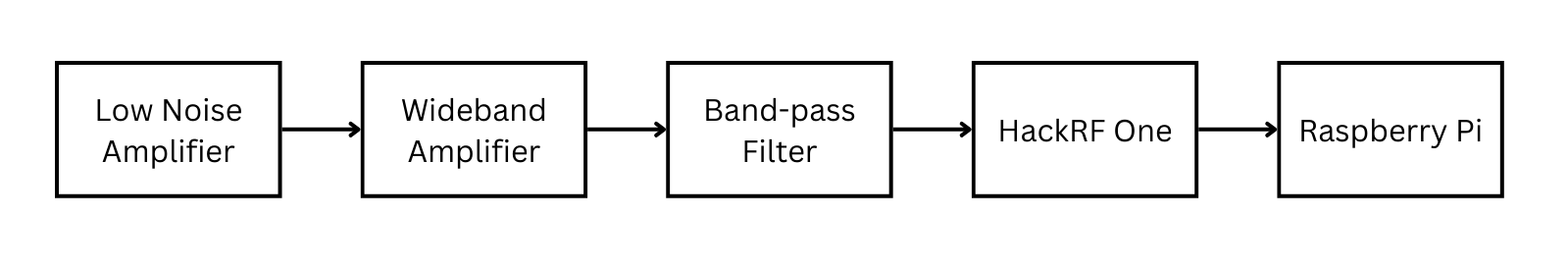}
    \caption{Block diagram of the RF signal chain.}
    \label{fig: RF_chain_block_diagram}
\end{figure*}

\begin{figure*}[ht!]
    \centering
    \includegraphics[width=\textwidth]{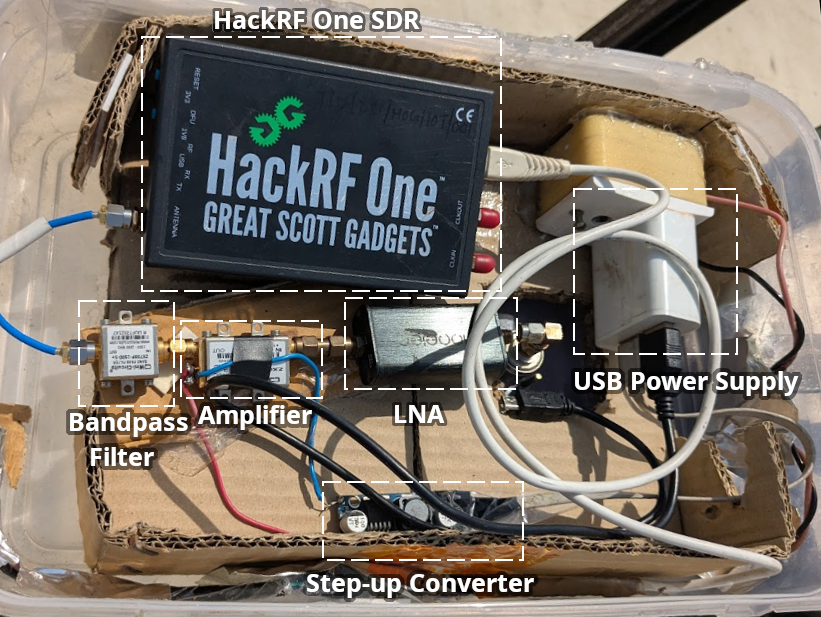}
    \caption{Image of the RF signal chain.}
    \label{fig: electronics}
\end{figure*}

The radio frequency (RF) front-end of the pyramidal horn telescope was designed to receive and amplify weak signals at the neutral hydrogen (\HI{}) line frequency of 1.42\,GHz. It consists of a low-noise amplifier (LNA), a wideband amplifier, and a band-pass filter. The conditioned RF signal was then digitised using a HackRF One software-defined radio (SDR) and subsequently processed and stored on a Raspberry Pi 5 single-board computer. The block diagram and image of the RF signal chain are shown in Figures~\ref{fig: RF_chain_block_diagram} and~\ref{fig: electronics}.

\subsection{Low-Noise Amplifier (LNA)}

The LNA used in the system was the Nooelec SAWbird + H1, a compact, self-contained module specifically designed for hydrogen line observations near 1.42\,GHz.  
It features a sharp band-pass filter with high out-of-band rejection and provides a typical gain of $\sim 40$\,dB within a 65\,MHz bandwidth centred at the target frequency.  
The module exhibits a low noise figure of $\sim 0.8$\,dB and operates with a nominal current draw of 120\,mA.  
The LNA was powered from one port of a dual-USB phone charger via a micro-USB cable.  
The SMA female connector of the LNA was connected to the N female connector in the probe via an N male–SMA male adapter and a right-angle SMA male–female connector.

\begin{figure}
    \centering
    \includegraphics[width=0.48\textwidth]{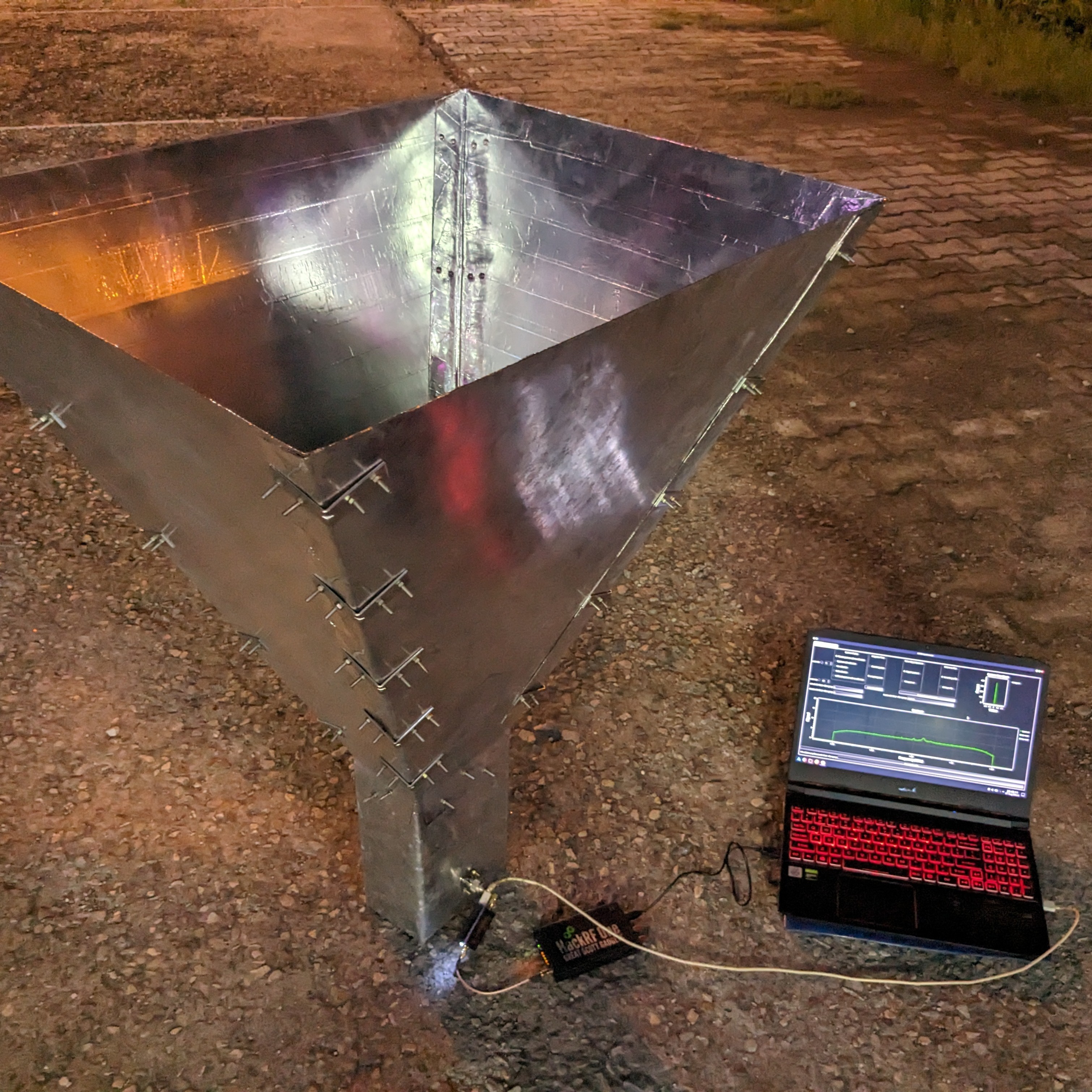}
    \caption{First light of the telescope on 7 August 2023.}
    \label{fig:first_light}
\end{figure}

The LNA alone provides sufficient gain for the detection of the \HI{} line without necessarily requiring any of the subsequent amplification and filtering stages. This is demonstrated in Figure~\ref{fig:first_light}, where the telescope achieved first light using only the LNA and the SDR. In the photograph, the horn antenna, LNA, SDR, and a laptop displaying the spectrum with the \HI{} line spike visible are shown during the first light test, conducted immediately after the horn was initially assembled, as the Galactic plane transited the local zenith.

\subsection{Wideband Amplifier}

Following the LNA, the signal was further amplified using the Mini-Circuits ZX60-1614LN-S wideband amplifier. This component offers a noise figure of $\sim 0.5$\,dB, a gain of $\sim 14$\,dB, and a third-order intercept point (IP3) of $+30$\,dBm at 1420\,MHz. It operates over the frequency range of 1217--1620\,MHz using a DC supply voltage between 11\,V and 13\,V, with a typical operating current of 42\,mA (maximum 50\,mA).
The second port of the aforementioned dual-USB phone charger powered the wideband amplifier, with the 5\,V and ground lines extracted from the USB connection and soldered to an XL6009 DC–DC step-up converter, whose output was adjusted to 13\,V using the onboard potentiometer and then connected to the amplifier.

\subsection{Band-Pass Filter}

A Mini-Circuits ZX75BP-1500-S+ lumped-element LC band-pass filter was employed to isolate the desired frequency band. This filter passes signals in the 1350\,MHz to 1650\,MHz range, attenuating out-of-band noise and interference while preserving the integrity of the hydrogen line signal.

\subsection{Software-Defined Radio Receiver}

Signal digitization was performed using the HackRF One, a versatile software-defined radio capable of both transmission and reception over a wide frequency range from 1\,MHz to 6\,GHz. This open-source, half-duplex transceiver supports a sampling rate of up to 20 million samples per second and provides 8-bit quadrature sampling (8-bit I and 8-bit Q). It is operated as a USB peripheral and is compatible with GNU Radio for signal processing and analysis.

\section{Simulation of Antenna Performance}
\label{Simulation of Antenna Performance}
\begin{figure}
    \centering
    \begin{subfigure}{0.48\textwidth}
        \centering
        \includegraphics[width=\linewidth]{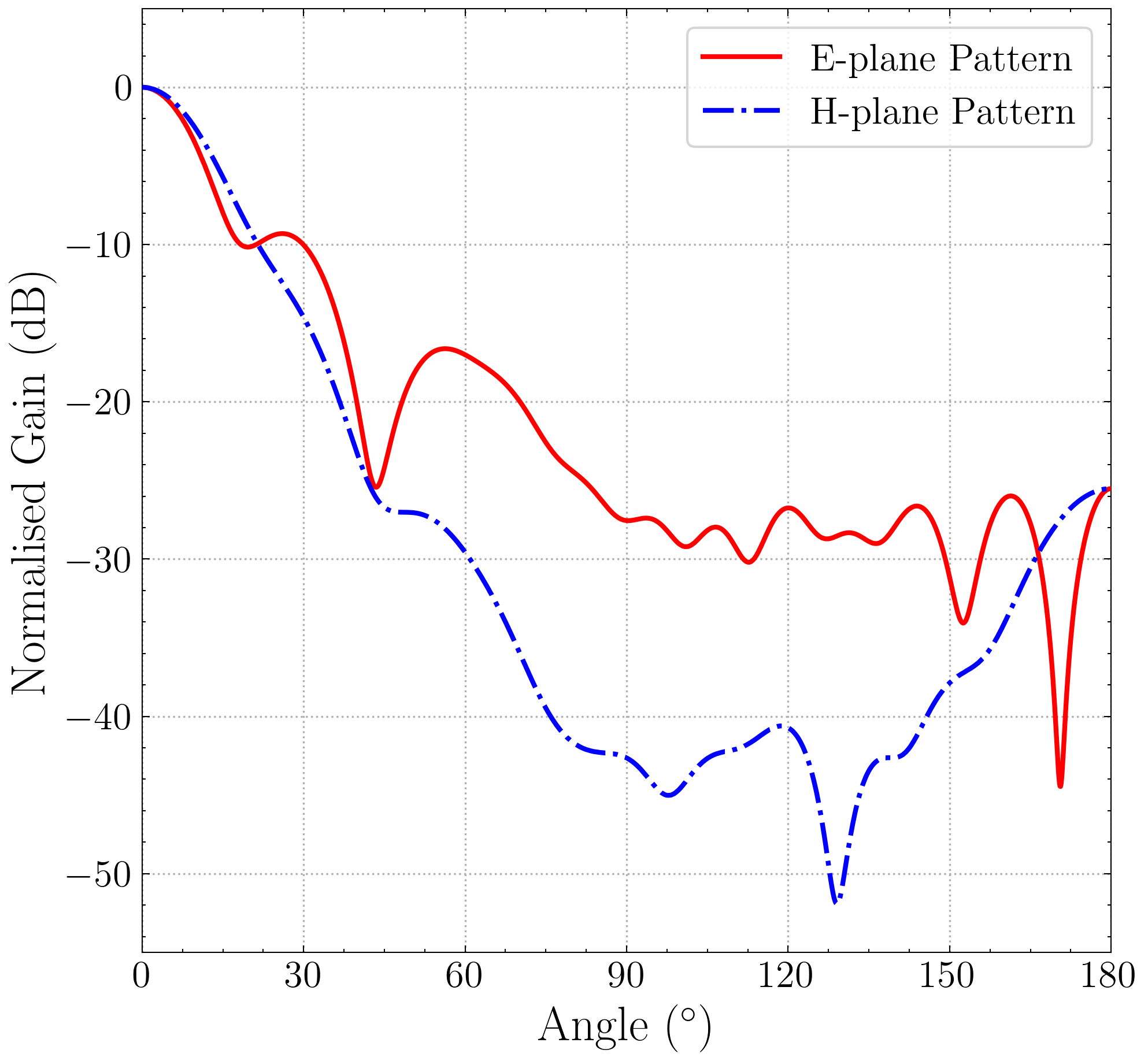}
        \caption{Normalised radiation patterns in the E-plane and H-plane.}
        \label{fig:2d_radiation_pattern}
        \end{subfigure}%
    \hfill
    \begin{subfigure}{0.48\textwidth}
        \centering
        \includegraphics[width=\linewidth]{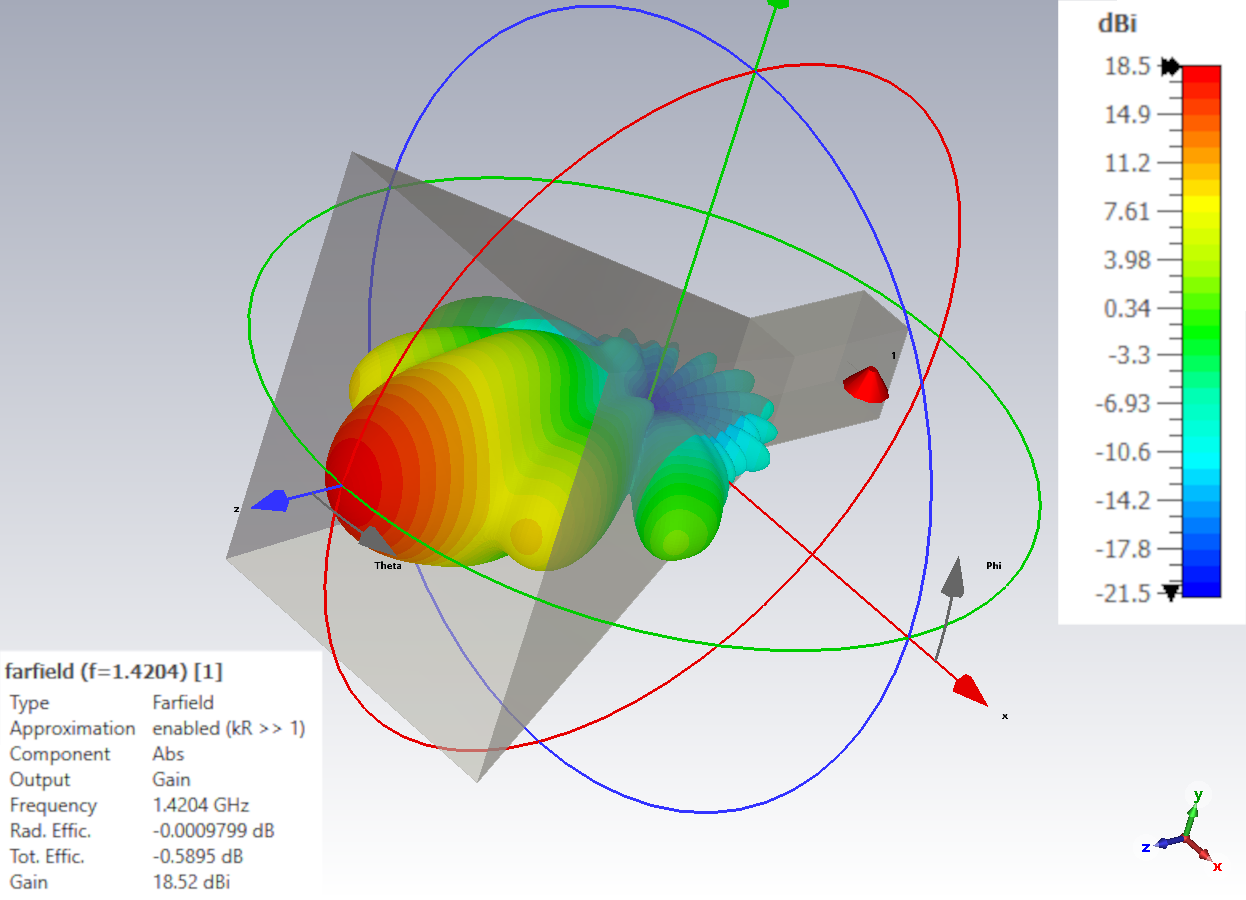}
        \caption{Three-dimensional simulated radiation pattern of the horn antenna.}
        \label{fig:3d_radiation_pattern}

    \end{subfigure}
    
        \caption{Simulated radiation characteristics of the horn antenna at 1420 MHz.}
    \label{fig:radiation_pattern}
\end{figure}

Electromagnetic (EM) simulations of the antenna were performed in CST Studio Suite (\citeauthor{cststudio2025}), a commercial EM analysis software package. The radiation pattern of the horn antenna was computed in the far-field (Fraunhofer) radiation zone. Figure~\ref{fig:radiation_pattern} shows the simulated two-dimensional and three-dimensional radiation patterns. The radiation pattern was analysed to determine the full-width at half-maximum (FWHM) of the main lobe, which was $18.1^\circ$ in the E-plane and $21.3^\circ$ in the H-plane. The simulated maximum gain is $18.5$~dBi, which was used to normalise the radiation patterns shown in Figure~\ref{fig:2d_radiation_pattern}.

The effective aperture area $A_\mathrm{e}$ for a gain ($G$) of $18.5$~dBi and wavelength ($\lambda$) is calculated as:

\begin{equation}
    A_\mathrm{e} = \frac{\lambda^{2} G}{4 \pi} = 0.251~\text{m}^2.
\end{equation}

The aperture efficiency $\eta$ is then obtained as:
\begin{equation}
    \eta = \frac{A_\mathrm{e}}{A_\mathrm{phys}} = 55.8\%,
\end{equation}
where $A_\mathrm{phys}$ is the physical aperture area of the horn. The total noise figure of the receiver chain, $NF_\mathrm{total}$, is calculated using the Friis formula (\citeauthor{1966raas.book.....K} \citeyear{1966raas.book.....K}):  
\begin{equation}
NF_{\text{total}} = NF_1 + \sum_{i=2}^{n} \frac{NF_i - 1}{\prod_{j=1}^{i-1} G_j},
\end{equation}
where $NF_i$ and $G_j$ are the noise figure and gain of the $i$-th and $j$-th stages respectively, and $n$ is the total number of stages.  

For the present setup, a total noise figure $NF_\mathrm{total} \sim 0.80004$\,dB is obtained using two terms of the formula. This corresponds to a noise temperature of approximately $60.68$\,K at a reference temperature of $300$\,K. It is well known that the overall noise figure is dominated by the first amplifying stage, which is also true in the present experimental setup, as evidenced by the first stage of the RF chain, the LNA, having a noise figure of $\sim 0.8$\,dB.

\section{Software Architecture}
\label{Software Architecture}

\subsection{GNU Radio Flowgraph Design}
\begin{figure*}
    \centering
    \includegraphics[width=\textwidth]{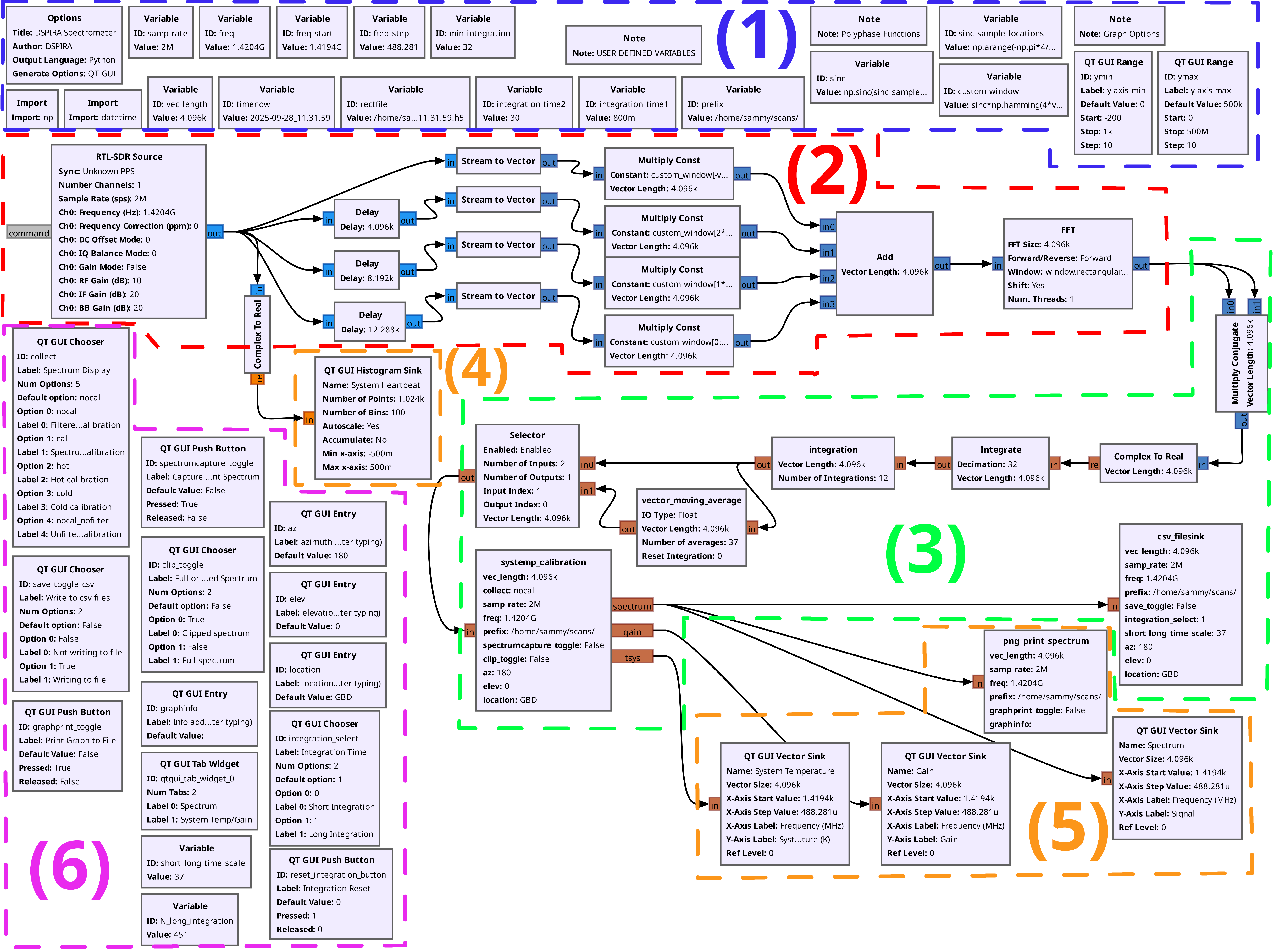}
    \caption{GNU Radio Flowchart. The blocks in the blue box (1) define usage options, imports, and variables. The blocks in the red box (2) show the RTL-SDR source, polyphase filter, and FFT calculation block. Following that, the blocks in the green box (3) perform integration, averaging, and file output. The blocks in the orange boxes (4 and 5) include all QT GUI display modules, while those in the pink box (6) manage the options for the GUI modules.}
    \label{fig: GNU Radio Flowchart}
\end{figure*}

The software spectrometer was implemented using a custom GNU Radio flowgraph, as shown in Figure~\ref{fig: GNU Radio Flowchart}, that performed a series of signal processing steps tailored for hydrogen line observations. The modular structure of the flowgraph allows real-time visualisation of system temperature, gain, and signal spectrum, as well as user control over integration time, frequency settings, and output modes.

\subsection{Fourier Transform Pipeline}
The HackRF SDR receives the RF signal centred around the 21 cm hydrogen line. The receiver frequency was set to 1419.5 MHz within GNU Radio. A stream-to-vector block segmented the incoming data stream into vectors of length 4096. To enable efficient spectral estimation and reduce spectral leakage, the data was passed through a polyphase filter bank that applies window functions with increasing delays. The outputs were combined and fed into a Fast Fourier Transform (FFT) block to compute the frequency spectrum.

To estimate power spectral density (PSD), the Fourier Transform output was multiplied by its complex conjugate. The resulting real-valued power spectrum was integrated over multiple windows, enhancing the signal-to-noise ratio. The final output was visualised in real-time using QT GUI sinks and was saved continuously for post-processing as well.

\subsection{Data Logging and Storage Format}
The processed spectra were timestamped and stored in CSV files using the Python \texttt{csv\_filesink} block within the GNU Radio environment. Each file name includes the acquisition time, allowing for easy temporal referencing. Calibration intervals: hot load measurements and drift scan segments are marked for later analysis.

The system was designed for unattended operation. When data logging is enabled, new files are automatically created after each integration interval. This automation ensures continuous data acquisition over long observation sessions.

\section{Observational Strategy}
\label{Observational Strategy}

\subsection{Scanning}
The antenna was aligned along the local meridian (North–South direction), allowing the centre of the beam to sweep across the sky from the northern to the southern horizon through the zenith when movement was not locked. At discrete elevation angles ranging from near the Celestial North Pole down to the southern horizon, the antenna was fixed, and the sky was allowed to drift past the beam over a full 24-hour period. This resulted in a series of drift scans at different declinations, enabling coverage of the entire visible sky over time.

\subsection{Calibration}
Between each change in elevation, i.e., after each complete drift scan, the antenna was manually pointed toward the ground for more than an hour. This position served as the hot load, assumed to have a brightness temperature of approximately 300 K. For cold load calibration, each drift scan was examined to identify regions with the highest Galactic latitude, i.e., areas farthest from the Galactic plane, where the contribution from neutral hydrogen is minimal. These regions predominantly contain only atmospheric emission, cosmic microwave background (CMB), and receiver noise, and are therefore used to define the cold load reference, assumed to have a brightness temperature of approximately 10 K.

The peak brightness of the \HI{} line in the collected spectra is significantly less than the peak brightness in the spectra obtained from radio telescopes with better spatial resolution, such as in the HI4PI all-sky database of Galactic \HI{} (\citeauthor{HI4PI} \citeyear{HI4PI}), which is based on the Effelsberg-Bonn \HI{} Survey (EBHIS) (\citeauthor{EBHIS} \citeyear{EBHIS}) and the Galactic All-Sky Survey (GASS) (\citeauthor{McClure-Griffiths_2009} \citeyear{McClure-Griffiths_2009}). This is because the temperature detected at an antenna $T_A$ is 
\begin{equation}
    T_A = \frac{\int_0^{2\pi} \int_0^{\pi} G(\theta, \phi) \, T_b(\theta, \phi) \, sin(\theta) \, d\theta \, d\phi} {\int_0^{2\pi} \int_0^{\pi} G(\theta, \phi) \, sin(\theta) \, d\theta \, d\phi},
\end{equation}
where $G(\theta, \phi)$ is the antenna gain pattern, and $T_b(\theta, \phi)$ is the brightness temperature of the sky. In plain words, the temperature $T_A$ is the weighted average of the temperature of the sky (and the ground) weighted by the beam pattern of the antenna. A beam pattern with a smaller spread will be truer to the actual nature of the sky, and conversely, a beam pattern with a larger spread will show a smudged view of the sky. The Effelsberg 100-m Radio Telescope, which was used in EBHIS, and the Murriyang, the 64-m CSIRO Parkes Radio Telescope, which was used in GASS, have enormously better beam patterns and sensitivities than our humble horn antenna, and observe a multitude of narrowly separated peaks in the \HI{} line spectrum near the Galactic plane, where we can only resolve a few widely separated peaks. 

We convoluted the spectra from the HI4PI database with the simulated beam pattern we discussed in Section~\ref{Simulation of Antenna Performance}. The peak brightness temperatures of the resulting beam--convolved spectra were compared with the peak temperatures measured in our crudely calibrated observations. This comparison was performed in $5^\circ$ bins in Equatorial coordinates. From the mean of the ratio of the HI4PI--based peak temperatures to the observed peak temperatures, we derived a calibration factor for each declination drift scan. These factors were subsequently applied as multiplicative corrections to the corresponding observed spectra. The difference between the positions of the peaks was used to correct a systematic frequency drift observed in the spectra, presumably instrumental in origin. The authors' best guess is that the drift is produced by the 25 MHz X1 crystal oscillator onboard the HackRF One that is used to generate the clock signal for both the local oscillator and ADC/DAC (see the device's block diagram in \citeauthor{HackRFArchitecture}, \citeyear{HackRFArchitecture}). However, this interpretation should be taken with a heap of salt, as the above specified part is rated to have a frequency stability of less than 20 ppm,  while the largest drift observed in our spectra was $\sim$ 40 ppm.

\section{Data Analysis}
\label{Data Analysis}

\subsection{RFI Detection and Mitigation}

Radio-frequency interference (RFI) appears in the spectra as narrow, high-amplitude spikes at specific frequencies. To suppress these contaminating features, we first applied a threshold-based flagging method. After flagging, we fit a sum of narrow Gaussian functions plus a smooth baseline to the affected regions. Each Gaussian was centred at a detected RFI spike, and its parameters (amplitude, width, centre) were optimised to model the shape of the interference. The fitted RFI model was then subtracted from the raw spectrum.

This multi-Gaussian approach was implemented using Python and enabled the restoration of underlying Galactic signals while effectively removing narrowband interference. The RFI mitigation was applied only for Galactic longitudes between 0° and 270° because, beyond this range, the signals from the \HI{} line were sufficiently redshifted to overlap with the RFI, which prevents effective filtering. 

\subsection{Galactic Mapping}

Following RFI mitigation, we averaged all calibrated spectra corresponding to each sky position, binned in Equatorial coordinates. Averaging multiple observations enhanced the signal-to-noise ratio and reduced random noise.

For each region, we fit the averaged spectrum using a sum of Gaussian profiles plus a constant baseline. The value of the sum of Gaussian profiles at the position of the Gaussian with the highest amplitude was used as the \HI{} signal strength at that location. The result was a map of neutral hydrogen intensity across the observed sky regions.

\subsection{Doppler Velocity Calculation}

The observed frequency shift of the \HI{} 21-cm line was converted into a Doppler velocity using the radio definition:

\begin{equation}
v = c \frac{\Delta f}{f_0},
\end{equation}

where \( f_0 = 1420.4058 \) MHz is the rest frequency, \( \Delta f = f_0 - f_{\text{obs}} \), and \( c \) is the speed of light. This yields the line-of-sight velocity relative to the observer's frame on Earth.

To convert these velocities to the Local Standard of Rest (LSR) frame, a two-step correction was applied:

\begin{enumerate}
  \item The barycentric velocity of Earth at the time of observation was calculated using the geographic location of the telescope and the observation time.
  \item The solar peculiar motion with respect to the LSR was calculated by adopting the values \((U, V, W) = (11.1, 12.24, 7.25)\) km/s, following the determination by \citeauthor{2010MNRAS.403.1829S} (\citeyear{2010MNRAS.403.1829S}), where, \(U\) is directed radially inwards toward the Galactic centre, \(V\) is along the direction of Galactic rotation, and \(W\) points vertically upwards toward the North Galactic Pole. 
\end{enumerate}

\begin{figure}
    \centering
    \includegraphics[width=0.48\textwidth]{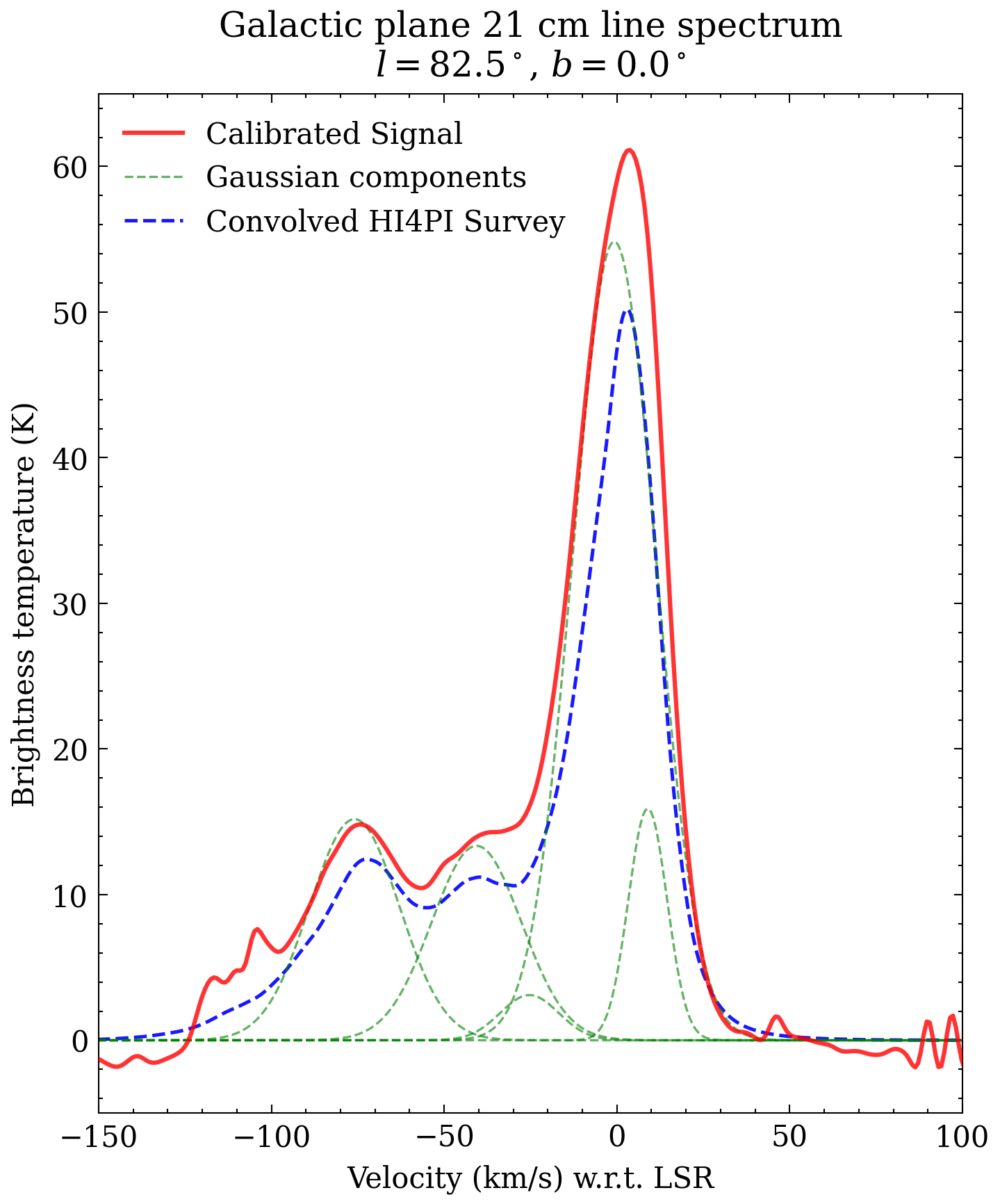}
    \caption{21\,cm \HI{} line spectrum at Galactic longitude $\ell=82.5^\circ$ and latitude $b=0^\circ$, approximately in the direction of the Cygnus constellation.}
    \label{fig:spectrum}
\end{figure}

Both the barycentric and solar peculiar velocities were projected onto the line of sight (LOS) towards the target source by taking the dot product of each velocity vector with the unit vector along the LOS. The sum of these two projections was taken as the velocity correction required, and was computed individually for each observation using the target's coordinates and the time of observation (see \citeauthor{fghigo_radvelcalc} \citeyear{fghigo_radvelcalc} for a web-utility to get the velocity correction for Green Bank Telescope, West Virginia). Figure~\ref{fig:spectrum} presents the 21\,cm \HI{} line spectrum obtained by averaging all spectra within a $5^\circ$ bin centred at $(\ell,b)=(82.5^\circ,0^\circ)$, which is roughly the direction of the Cygnus constellation. The Gaussian profiles fitted to the spectrum and the beam--convolved HI4PI spectrum at that region are also shown.

\subsection{Velocity–Radius Curve Estimation}
\label{Velocity–Radius Curve Estimation}

To estimate the Galactic rotation curve, we employed the tangent-point method within the inner Galaxy range \( 0^\circ < \ell < 90^\circ \). For a given Galactic longitude \( \ell \), the tangent point is the location along the line of sight that is closest to the Galactic centre, where the line is tangent to a circular orbit (the assumption that orbits are circular is maintained throughout this study).

From Figure~\ref{fig:geometry}, it is clear that from the perspective of an observer in the LSR frame, orbiting the Galactic centre with velocity $V_0$ at a distance $R_0$ away from it, a target that is orbiting the Galactic centre with velocity $V(R)$ at a distance $R$ away from it, in the Galactic plane at a Galactic longitude $\ell$, moves away with velocity $V_{\text{obs}}$:

\begin{equation}
\label{eq:velocity_obs}
    V_{\text{obs}} = V(R) \sin(\theta+\ell) - V_0 \sin \ell.
\end{equation}

Clearly, along a particular line-of-sight, the maximum line-of-sight velocity \( V_{\text{max}} \) is observed from the clouds of hydrogen that move satisfying $\sin(\theta+\ell) = 1$, i.e. are on the tangent point for that the Galactic longitude of that line-of-sight, under a not-too-unreasonable assumption that $V(R)$ is not heavily varying with R. Thus, from the spectra from the Galactic plane, for each longitude $\ell$, we can calculate the corresponding Galactocentric radius and rotation velocity by selecting the most red-shifted Gaussian component using:

\begin{equation}
R = R_0 \sin \ell,
\quad
V(R) = V_{\text{max}} + V_0 \sin \ell
\end{equation}

where $V_{\text{max}}$ is the observed velocity of the most red-shifted Gaussian component.

The uncertainty in \( V(R) \) was estimated from the standard deviation of the Gaussian fit (i.e., its width), while the uncertainty in \( R \) was derived from the longitude bin width. These calculations were implemented in Python by binning the data in Galactic longitude, extracting \( V_{\text{max}} \) from the Gaussian fits, and computing the rotation curve point-by-point.

\subsection{Spiral Arm Structure Derivation}

\begin{figure}
    \centering
    \includegraphics[width=0.48\textwidth]{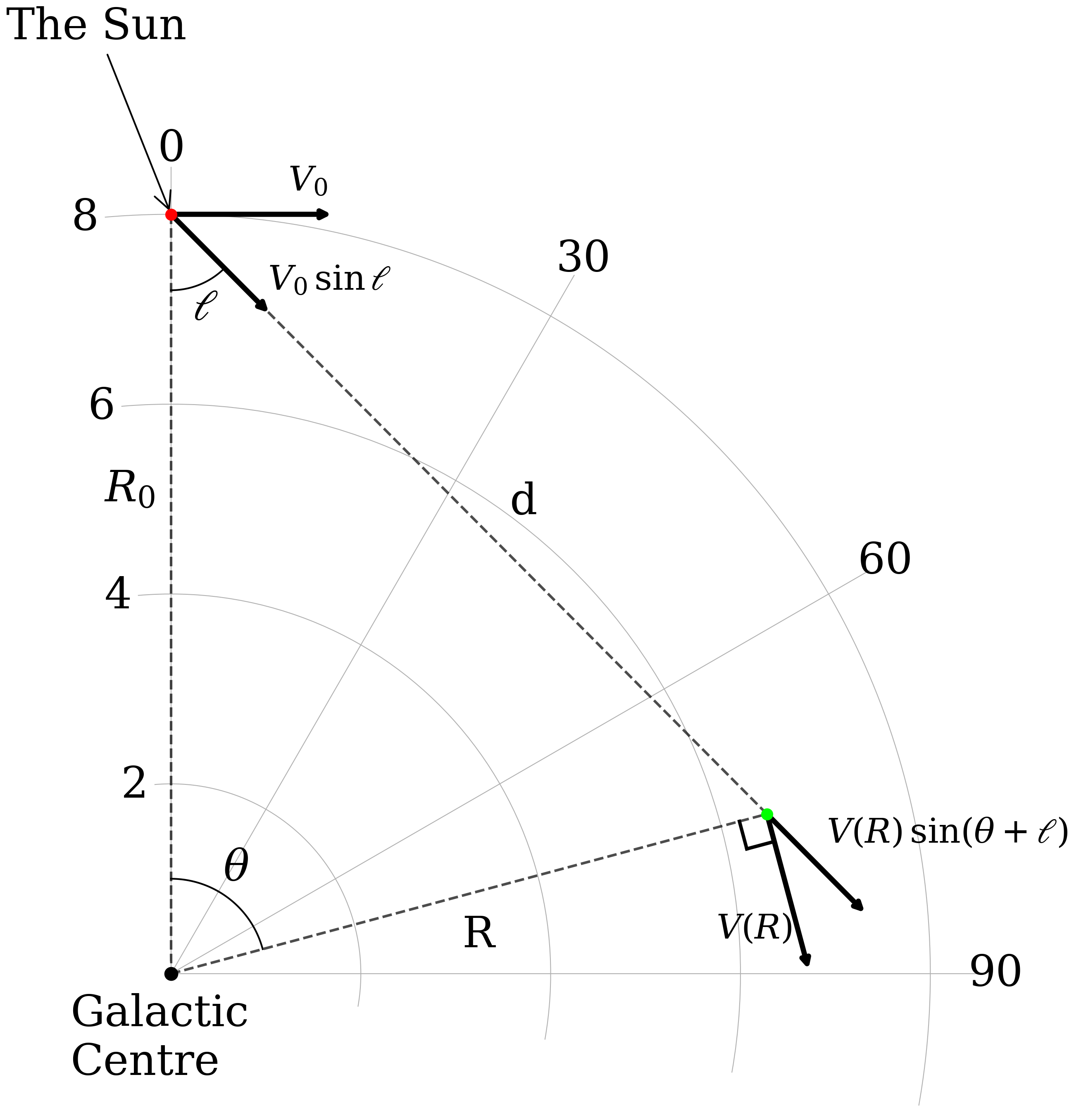}
    \caption{Schematic view of geometry of Galactic rotation and line of sight velocities in the Galactic plane.}
    \label{fig:geometry}
\end{figure}

To infer the large-scale spiral structure of the Milky Way, we make use the measured line-of-sight velocities of all the Gaussian components, not just the most red-shifted one. Using the law of sines and Figure~\ref{fig:geometry}, we get:

\begin{equation}
\frac{R_0}{\sin(\theta+\ell)} = \frac{R}{\sin \ell} = \frac{d}{\sin \theta},
\end{equation}
which, using Eq.~\eqref{eq:velocity_obs}, can be simplified to:
\begin{equation}
\theta = \pi - (\ell + \sin^{-1}(\frac{R_0}{R}\sin\ell)), 
\end{equation}
\begin{equation}
R = \frac{V(R)\,R \, \sin{\ell}}{V_0 \sin{\ell}+ V_{\text{obs}}}.
\end{equation}

For each Gaussian component at each Galactic longitude, the corresponding Galactocentric coordinate $(R, \,\theta)$ was determined under the assumption of a flat Galactic rotation curve with constant circular velocity $V(R) = V_0$.  

\section{Results and Discussion}
\label{Results and Discussion}

\subsection{Sky Maps}

\begin{figure}[h]
    \centering
    \begin{subfigure}{0.48\textwidth}
        \centering
        \includegraphics[width=\linewidth]{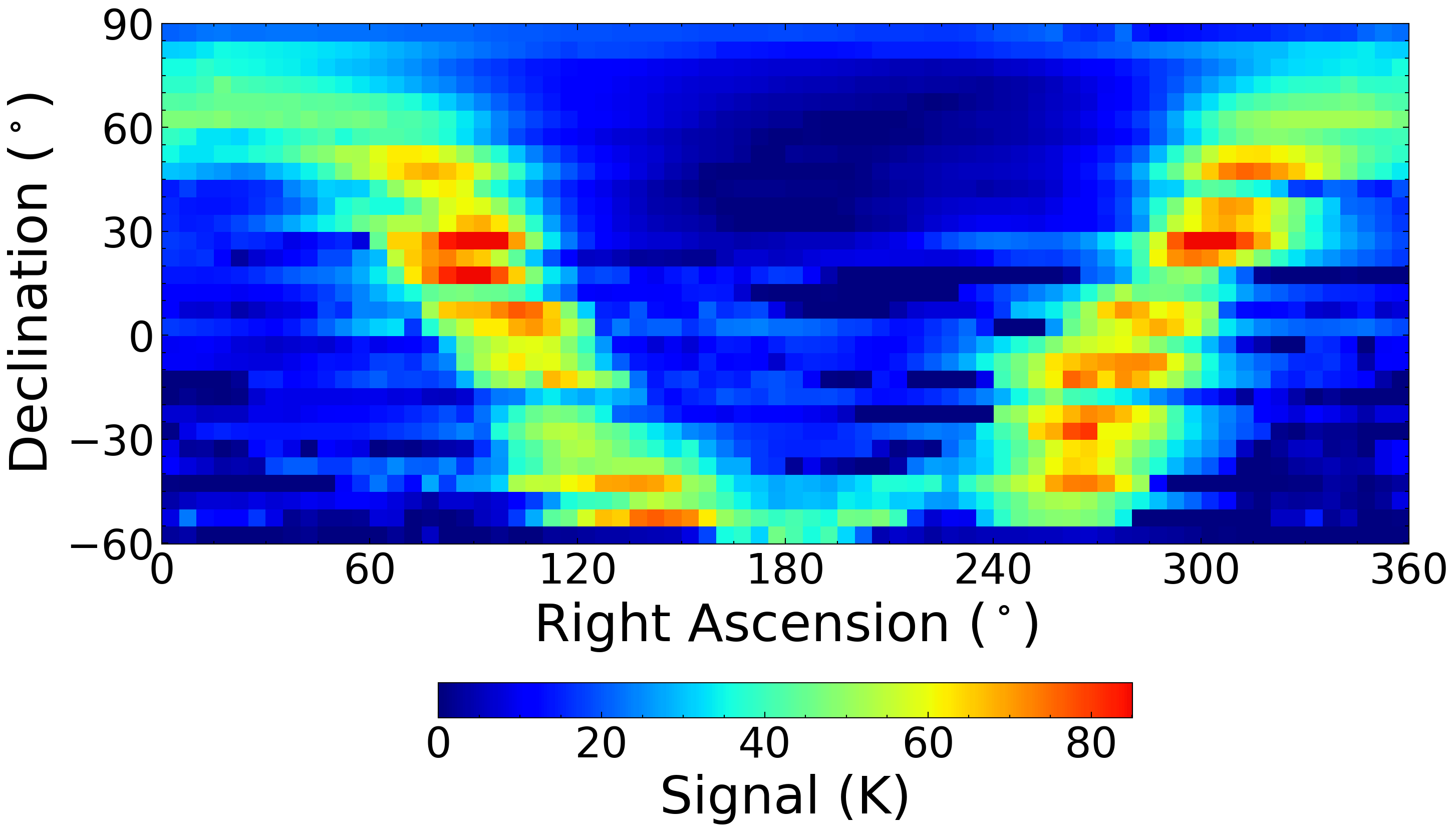}
        \caption{Heatmap in Equatorial coordinate system.}
        \label{fig:Galactic_heatmap_ra_dec}
    \end{subfigure}

    \vspace{0.5cm} 

    \begin{subfigure}{0.48\textwidth}
        \centering
        \includegraphics[width=\linewidth]{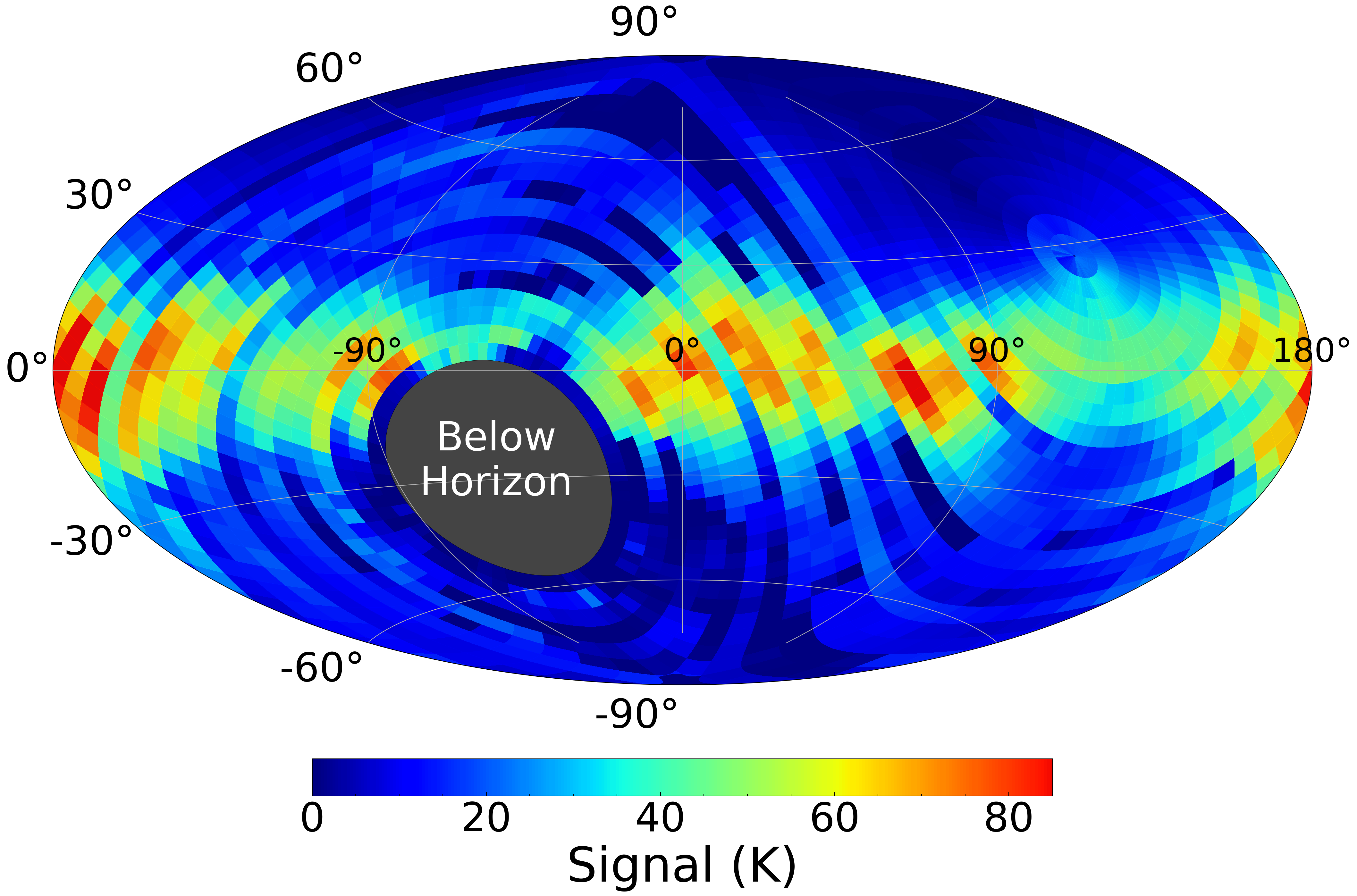}
        \caption{Heatmap in Galactic coordinate system.}
        \label{fig:Galactic_heatmap_aitoff}
    \end{subfigure}

    \caption{Comparison of Galactic heatmaps: (a) in Equatorial coordinates and (b) in Aitoff projection of Galactic coordinates.}
    \label{fig:Galactic_heatmaps_combined}
\end{figure}

After calibrating and processing the spectral data, the sky maps in Figure~\ref{fig:Galactic_heatmaps_combined} were generated to visualise the intensity of the 21 cm line signal at different points in the sky.

Figure~\ref{fig:Galactic_heatmap_ra_dec} was created using Equatorial coordinates and extends to a minimum declination of approximately $-60^{\circ}$, corresponding to the southernmost sky visible from the observing site at Roorkee, India ($\varphi \sim29.86^{\circ}$ N). Figure~\ref{fig:Galactic_heatmap_aitoff} used Galactic coordinates projected with an Aitoff projection and includes a grey region around the southern celestial pole that remains permanently below the horizon. These maps trace the structure of the Galactic plane, where the \HI{} signal is strongest, and show a decrease in intensity at higher Galactic latitudes, as expected.

\subsection{Rotation Curve}
\begin{figure*}
    \centering
    \includegraphics[width=0.9\textwidth]{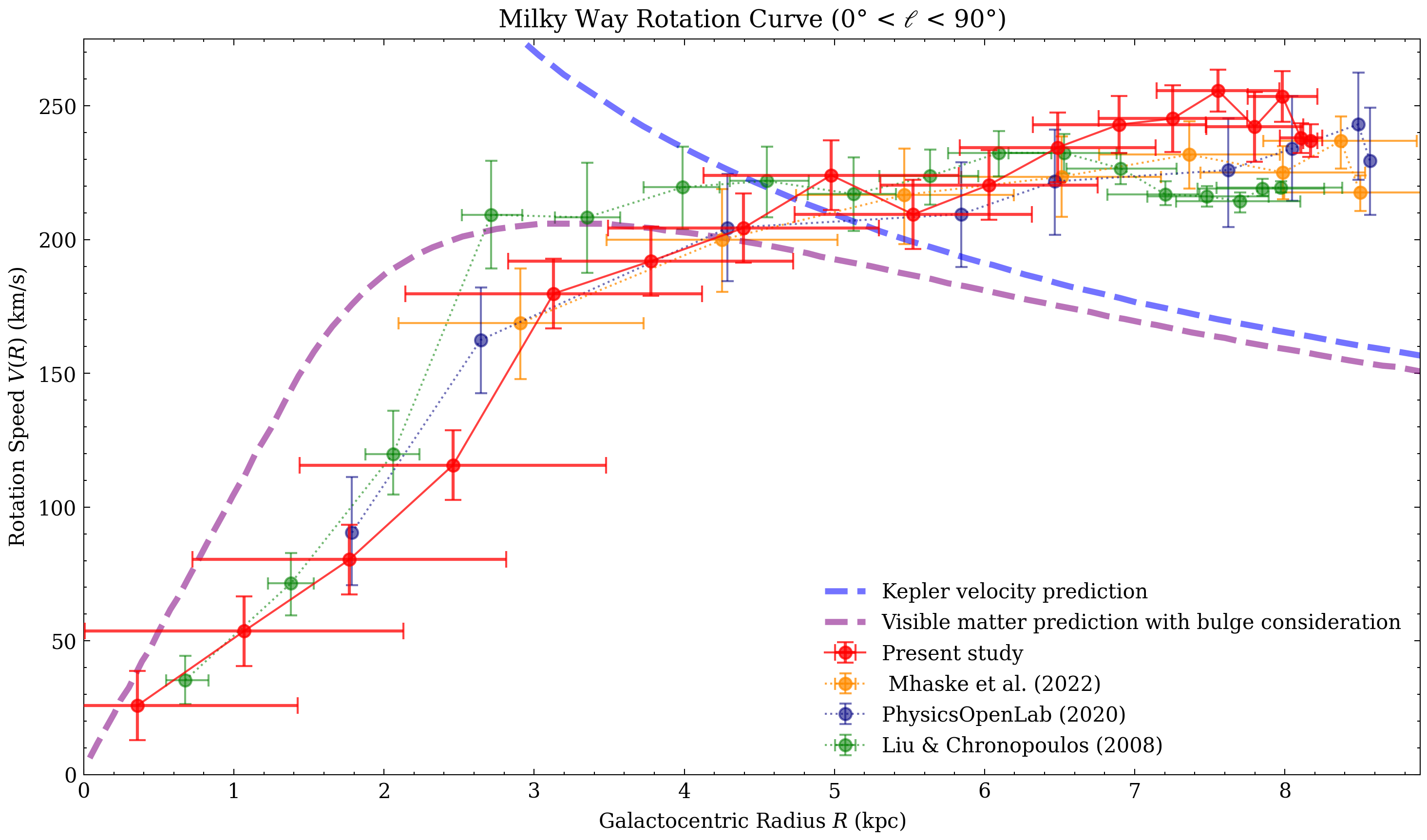}
    \caption{The Velocity-Radius rotation curve for the Milky Way galaxy.}
    \label{fig:rotation_curve}
\end{figure*}

The Doppler shifts observed in the \HI{} spectral lines were used to estimate the rotation curve of the Milky Way. Using the tangent point method, as described in Section~\ref{Velocity–Radius Curve Estimation}, for longitudes between 0° and 90°, the maximum line-of-sight velocities were extracted and used to calculate the circular velocity of Galactic rotation at different radii from the Galactic centre. Due to the geometric limitation of the tangent-point method being applicable only when the line-of-sight is tangential to an orbit, the rotation curve derived from this method ends at $R_0$. 

A couple of previous studies have attempted to extend the Galactic rotation curve beyond $R_0$ using an alternative approach based on a so-called \enquote{velocity-vector method} and the application of Oort constants (\citeauthor{Pandian2022-pd} \citeyear{Pandian2022-pd}, \citeauthor{Makhijani2024} \citeyear{Makhijani2024}). However, such a method is not justified because the Oort constants are defined through a first-order Taylor expansion of the Galactic angular velocity about the Solar neighbourhood and are strictly valid only for tracers located at distances much smaller than $R_0$ $(d \ll R_0)$. Consequently, its use for inferring the rotation curve at Galactocentric radii much larger than $R_0$ is not supported by the underlying assumptions. We therefore restrict our analysis to radii $R \leq R_0$, where the tangent-point method is applicable. The rotation curve can be and has been extended to $R > R_0$ using other sources like stars and masers, and more sophisticated methods, including VLBI and spectroscopy (\citeauthor{sofue2020rotationcurvemilkyway} \citeyear{sofue2020rotationcurvemilkyway}).

The resulting rotation curve, shown in Figure~\ref{fig:rotation_curve}, is consistent with the general trend reported in the literature (\citeauthor{liu2008hydrogen} \citeyear{liu2008hydrogen}, \citeauthor{Mhaske2022} \citeyear{Mhaske2022}, \citeauthor{physicsopenlab} \citeyear{physicsopenlab}). The curve rises steeply near the Galactic centre, reflecting the increasing enclosed mass at small radii, and then gradually flattens at larger distances from the centre. This deviates from the Keplerian and visible matter predictions, according to which the velocity of circularly orbiting matter should decrease as $V \propto 1/ \sqrt{R}$ (\citeauthor{liu2008hydrogen} \citeyear{liu2008hydrogen}).  This flatness supports the presence of a halo of dark matter, which is invisible in optical and radio frequencies, but provides the additional gravitational force needed to maintain higher-than-expected rotation speeds at large distances (\citeauthor{Corbelli_2000} \citeyear{Corbelli_2000}).

\begin{figure*}[ht!]
    \centering
    \includegraphics[width=0.85\textwidth]{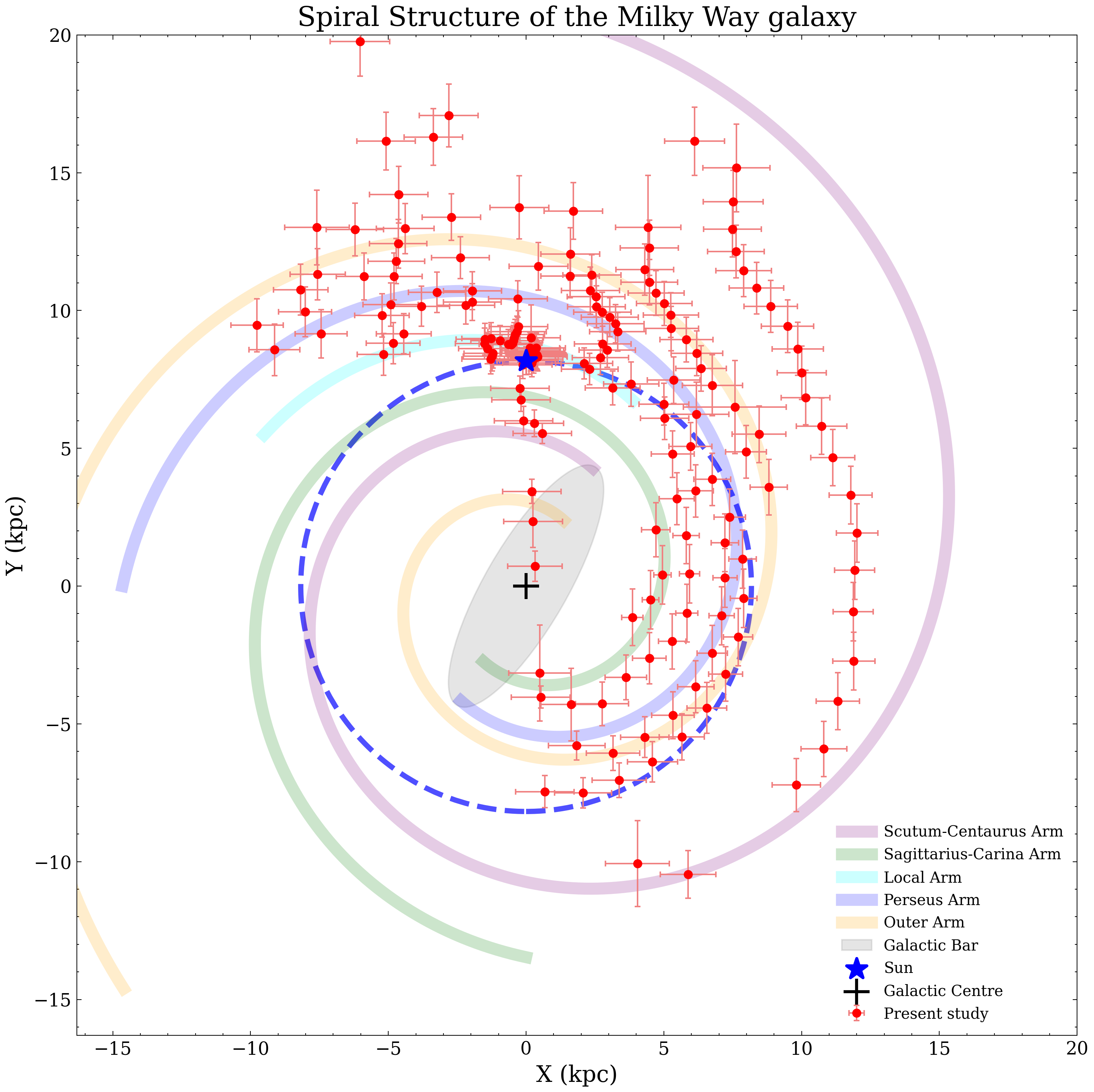}
    \caption{The structure of the Milky Way galaxy.}
    \label{fig:spiral_structure}
\end{figure*}

\subsection{Spiral Arm Structure}

Figure~\ref{fig:spiral_structure} shows the distribution of hydrogen by mapping the spiral arm tracers of \HI{} in the Galactic plane. The Sun and Galactic centre are marked as references, with the solar circle ($R=R_0$), the schematic trajectories of the major Milky Way spiral arms (\citeauthor{10.1093/mnras/stad3350} \citeyear{10.1093/mnras/stad3350}) and the Galactic bar (\citeauthor{10.1093/mnras/stv745} \citeyear{10.1093/mnras/stv745}) overlaid for context. The spiral nature of the galaxy can be seen clearly, and the tracer positions are consistent with the locations of the spiral arms up to the Outer Arm. The Scutum-Centaurus arm, however, appears much closer to us than expected.

\section{Conclusion}

The construction and working of the pyramidal horn radio telescope to detect the \HI{} line is presented in this work. Our pyramidal horn radio telescope successfully detected the Galactic 21 cm \HI{} line and produced spectra suitable for mapping the hydrogen distribution and deriving the Milky Way rotation curve and the spiral arm structure. Despite limitations imposed by the sensitivity, beamwidth, and RFI contamination, the results reproduce the expected rotation curve and reveal large-scale Galactic structure, demonstrating the effectiveness of this low-cost design and methodology.

Future work in the Radio Telescope Project of the Physics and Astronomy Club will focus on advancing the observational and instrumental capabilities. Building on earlier work in which a Ku-band TV dish was motorised into a small radio telescope (\citeauthor{Bhattacharjee:2024jvi} \citeyear{Bhattacharjee:2024jvi}), we intend to replace its feed with one capable of observing the hydrogen line and equip it with the same electronics and software as the current horn system. Operating multiple such antennas in interferometric mode opens up new possibilities for higher-resolution studies. In parallel, efforts are underway to construct a 5-meter radio dish, which in its initial stage will operate in a fixed zenith mode for hydrogen-line observations. This dish will subsequently be motorised to enable full-sky scanning and upgraded with higher-sensitivity feeds and backends, ultimately enabling pulsar observations and timing studies  (\citeauthor{Grover2024} \citeyear{Grover2024}).

All data products and analysis codes used in this study are publicly available on \href{https://github.com/CtrlShiftSammy/mrt}{Github}. 

\section{Acknowledgments}

 We thank Bhal Chandra Joshi (NCRA--TIFR) for generously sharing his expertise and providing guidance in technical challenges. We are especially grateful to T. Prabu (RRI) for his extensive support, including contributions to several sections of this work. Their feedback was instrumental in shaping the direction of this research. KB extends his gratitude to his peers Ayush Ashray Nishad, Krish Shah, and Shivam Singh Aswal for their assistance in constructing much of the experimental setup. The authors gratefully acknowledge Mr. Javed Shah of Swastik Tins Pvt. Ltd. for generously providing the can used as the waveguide in this study free of charge. The authors thank the technical and administrative staff of the Department of Physics, the Student Technical Council, and the Tinkering Lab at IIT Roorkee for providing access to facilities and equipment.  \vspace{-1em}

\bibliographystyle{plainnat}
\bibliography{main}

\end{document}